\documentclass[12pt]{iopart}
\usepackage{iopams,epsfig}
\newcommand{\text}{\mbox}
\newtheorem{lemma}{Lemma}

\newtheorem{theorem}{Theorem}

\begin{document}

\title[Localized magnon states]{Linear independence of localized magnon states}

\author{Heinz~-~J\"urgen Schmidt$^1$
\footnote{To whom correspondence should be addressed
(hschmidt@uos.de)},
Johannes Richter$^2$
and Roderich Moessner$^3$}

\address{$^1$  Fachbereich Physik, Universit\"at Osnabr\"uck,
Barbarastr. 7, 49069 Osnabr\"uck, Germany}
\address{$^2$ Institute of Theoretical Physics, University of Magdeburg,
P.~O.~Box 4120, D-39016 Magdeburg, Germany}
\address{$^3$ Laboratoire de Physique Th\'eorique, CNRS-UMR 8549,
Ecole Normale Sup\'erieure, Paris, France}

\begin{abstract}
  {At the magnetic saturation field,
certain frustrated lattices have a class of states known as ``localized
    multi-magnon states" as exact ground states.
The number of these states scales
exponentially with the number $N$ of spins and hence they have a finite
entropy also in the thermodynamic limit $N\rightarrow \infty$ provided they
    are sufficiently
linearly independent. In this article we present rigorous results
concerning the linear dependence or independence of localized
magnon states and investigate special examples. For large classes
of spin lattices including what we called the orthogonal type and
the isolated type as well as the kagom\'{e}, the checkerboard and
the star lattice we have proven linear independence of all
localized multi-magnon states. On the other hand the pyrochlore
lattice provides an example of a spin lattice having localized
multi-magnon states with considerable linear dependence.}
\end{abstract}


\maketitle

\section{Introduction and summary\label{sec:I}}
Some years ago a class of exact ground states of certain frustrated spin systems was
discovered \cite{SSRS}.
These states can be characterized as localized one-magnon states ($1$-LM states) or
multi-magnon states ($a$-LM states, where $a$ denotes the number
of magnons involved) and span highly degenerate subspaces of
the space of ground states for a
magnetic field $h$ attaining its saturation value $h_{\text{\scriptsize sat}}$.\\
The $1$-LM states are superpositions of spin flips localized on a
zero-dimensional subsystem, called ``unit".
These units are not unique but always chosen to be
as small as possible. The reason for this is the desire to
obtain maximally independent or unconnected units which will host
collections of localized magnons without interaction, the so-called $a$-LM states.
\\

The LM states yield several spectacular effects near saturation
field: Due to these states at zero temperature there is a
macroscopic magnetization jump to saturation in spin systems
hosting LM states \cite{SSRS}. Furthermore, one observes a
magnetic field induced spin-Peierls instability \cite{RDJ}, and,
last but not least, a residual ground-state entropy at the
saturation field \cite{RSH,ZT,ZH,DR,ZT05,DR06}. Whereas huge
ground-state manifolds as such are not unusual in frustrated
magnetism \cite{RMcjp}, exact degeneracies in {\em quantum}
frustrated magnets are not so common. This ground-state entropy
leads to interesting low-temperature properties at $h\approx
h_{\text{\scriptsize sat}}$ such as an enhanced magnetocaloric
effect or an extra maximum in the
 specific heat at low temperatures.
By mapping
the spin model onto solvable
hard core models of statistical mechanics  exact analytical expressions
for the contribution
of LM states to the low-temperature thermodynamics
can be found
\cite{ZT,ZH,DR,ZT05,DR06}.

In the context of these calculations the following questions
arise.
\begin{description}
\item[]$\quad$ Does the number of LM states equal the dimension of the true ground state space for
$h = h_{\text{\scriptsize sat}}$,
at least in the thermodynamic limit $N\rightarrow\infty$?
\item[]In particular:
\item[(a)] Is the set of LM states linearly independent, and
\item[(b)] Does the set of LM states span the whole subspace of ground states or are there more ground states
than those of LM type?
\end{description}
The answers to these questions are, on the one hand, of general
interest, since they provide exact statements about non-trivial
quantum many body systems, but are, on the other hand, crucial for
the contribution of the LM states to the thermodynamics at low
temperatures and magnetic fields close to the saturation field.
However, so far these questions have not been considered in the
corresponding publications \cite{RSH,ZT,ZH,DR,ZT05,DR06}, except
in Ref.~\cite{ZT05} where some aspects for the sawtooth chain and
the kagom\'{e} lattice are briefly discussed.
In this paper we address mainly question
(a) concerning linear independence. We group frustrated lattices
into classes, for three of which we show {\em rigorously} that the
multi-magnon states are indeed linearly independent; and, for a fourth
class, we  provide an example
that this is not generally the case. Question (b) concerning non-LM
ground states will be treated elsewhere.
The four classes, explained in detail below, are the following:\\

\begin{figure}
\begin{center}
\includegraphics[width=70mm]{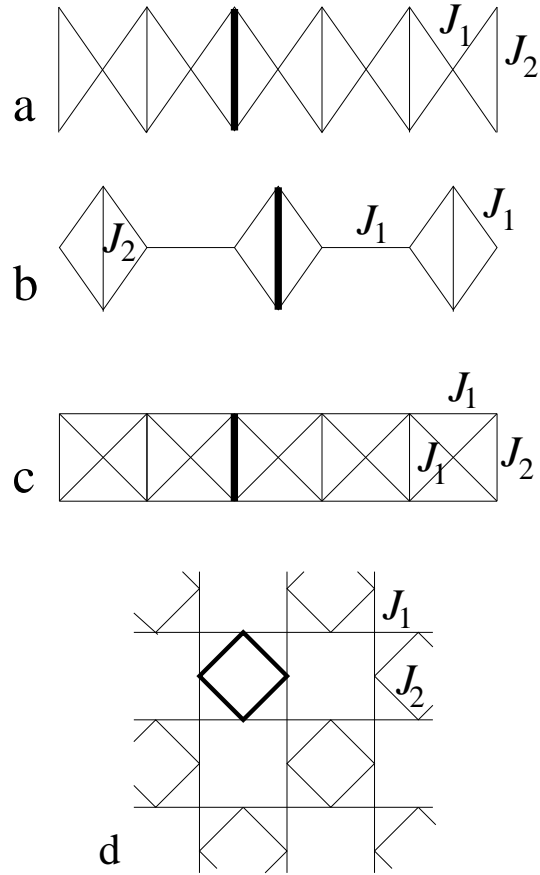}
\caption{\label{fig01}Examples of spin lattices admitting only
orthogonal LM states (orthogonal type): (a) the diamond chain
\cite{14}, (b) the dimer-plaquette chain \cite{15}, (c) the
frustrated ladder \cite{16} (where the spins are sitting only on
the squares, not on the intersections of the diagonals) and (d)
the square-kagom\'{e} lattice \cite{17}. Note that the ratios of
the exchange constants $J_1, J_2$ have to assume certain values in
order to admit LM states, for more details see Ref.~\cite{DR06}.
$1$-LM states are localized on small subsystems (``units")
typically indicated by thick lines. }
\end{center}
\end{figure}

\begin{figure}
\begin{center}
\includegraphics[width=70mm]{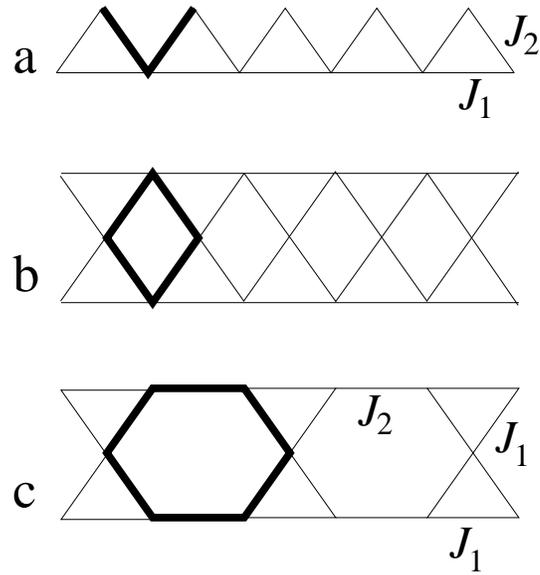}
\caption{\label{fig02}Examples of spin lattices admitting only LM states with isolated sites (isolated type):
(a) the sawtooth chain \cite{18}, (b) the kagom\'{e} chain I \cite{19}
and (c) the kagom\'{e} chain II \cite{20}.
Note that the ratios of the exchange constants $J_1, J_2$ in the cases (a) and (c)
have to assume certain values in order to admit LM states. In the case (b) all exchange constants are positive and equal.
$1$-LM states are localized on small subsystems (``units") typically indicated by thick lines.
}
\end{center}
\end{figure}

\begin{figure}
\includegraphics[width=\columnwidth]{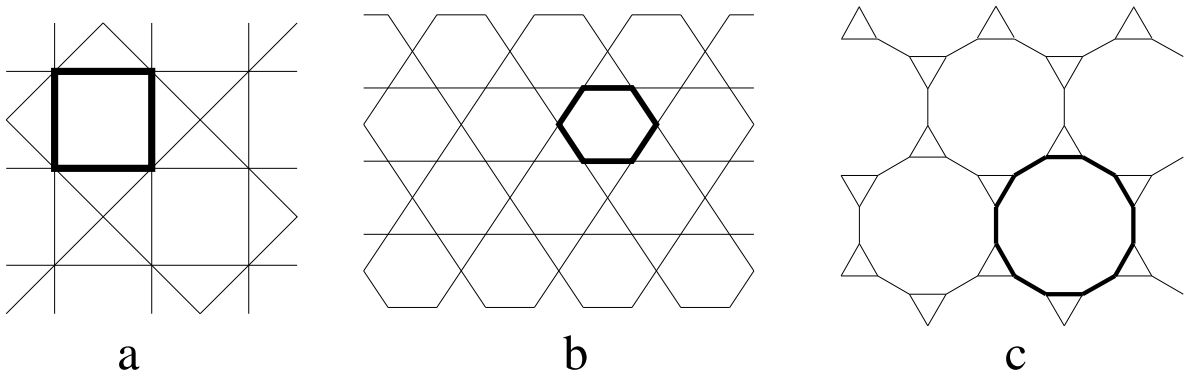}
\caption{\label{fig03}Examples of spin lattices admitting only $1$-LM states satisfying exactly one linear relation
(codimension one type):
(a) the checkerboard lattice or two-dimensional pyrochlore \cite{21}
(where the spins are sitting only on the squares, not on the intersections of the diagonals),
(b) the kagom\'{e} lattice \cite{22,RSH}, and
(c) the star lattice \cite{RSH,23}. All exchange constants are positive and equal.
$1$-LM states are localized on small subsystems (``units") typically indicated by thick lines.
}
\end{figure}

\begin{figure}
\includegraphics[width=\columnwidth]{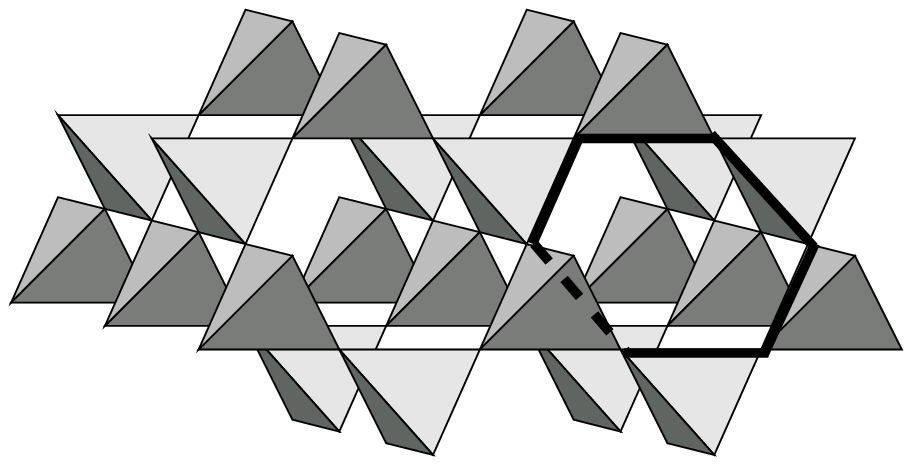}
\caption{\label{fig04}The pyrochlore lattice \cite{24} as an example of a spin lattice admitting $a$-LM states satisfying
more than one linear relation
(higher codimension type). All exchange constants are positive and equal.
$1$-LM states are localized on small hexagonal subsystems (``units") typically indicated by thick lines.
}
\end{figure}

\begin{description}
\item[1] Orthogonal type
\begin{description}
\item[(a)] Diamond chain (fig. \ref{fig01}a)
\item[(b)] Dimer-plaquette chain (fig. \ref{fig01}b)
\item[(c)] Frustrated ladder (fig. \ref{fig01}c)
\item[(d)] Square-kagom\'{e} (fig. \ref{fig01}d)
\end{description}
\item[2] Isolated type
\begin{description}
\item[(a)] Sawtooth chain (fig. \ref{fig02}a)
\item[(b)] Kagom\'{e} chain I (fig. \ref{fig02}b)
\item[(c)] Kagom\'{e}  chain II (fig. \ref{fig02}c)
\end{description}
\item[3] Codimension one type
\begin{description}
\item[(a)] Checkerboard lattice (fig. \ref{fig03}a)
\item[(b)] Kagom\'{e} lattice (fig. \ref{fig03}b)
\item[(c)] Star lattice (fig. \ref{fig03}c)
\end{description}
\item[4] Higher codimension type
\begin{description}
\item[(a)] Pyrochlore lattice (fig. \ref{fig04})
\end{description}
\end{description}

It turns out that the spin systems of orthogonal or isolated type always admit linearly independent
LM states. We will show that the lattices of class $3$ possess,
up to a factor,
exactly one linear relation between their
$1$-LM states (hence the wording ``codimension one"), but have linearly independent
$a$-LM states for $a>1$,
whereas the lattices of class $4$ have more than one linear relation between their
LM states (higher codimension).\\

In section $2$ we recapitulate the pertinent definitions for LM
states. Section $3$ is devoted to some general algebraic methods
and results related to linear independence. An elementary but
important tool is the ``Gram matrix" $G^{(a)}$ of all possible
scalar products between $a$-LM states, since linear independence
of the $a$-LM states is equivalent to $G^{(a)}$ being
non-degenerate, i.~e.~$\det G^{(a)}\neq 0$. It turns out that if
$G^{(1)}$ is non-degenerate then all $G^{(a)}$ will also be
non-degenerate for $a\ge 1$. These methods will be applied in
section $5$ to the special cases enumerated above. The main result
(theorem \ref{T5}) is contained in section $4$ and states that for
a class of codimension one lattices, including the checkerboard
(figure \ref{fig03}a), the kagom\'{e} (figure \ref{fig03}b), and
the star lattice (figure \ref{fig03}c), all $a$-LM states are
linearly independent for $a>1$, although the corresponding $1$-LM
states have codimension one. This result has been checked for some
finite versions of the checkerboard lattice and some values of
$a$, see table 1. We also confirmed theorem \ref{T5} for some
finite kagom\'{e} lattices but will not give the details here. For
the higher codimension case we have no general theorem but some
computer-algebraic and numerical results for certain finite
pyrochlore lattices (table 2). However, one can argue that the
codimension of $a$-LM states will be positive
for some $a>1$ due to localized linear dependencies between $1$-LM states,
see also \cite{HH} and \cite{Z2}.\\

Summarizing, we want to stress that the question of linear
independence of localized multi-magnon states is relevant for
thermodynamical calculations but, in general, non-trivial. For
certain mainly one-dimensional lattices including what we called
the orthogonal and the isolated types the problem can be solved by
relatively elementary considerations. Here all multi-magnon states
are linearly independent. The checkerboard, the kagom\'{e} and the
star lattice are more involved, but we have shown that they
possess linearly independent localized multi-magnon states except
for $a=1$. For the pyrochlore lattice we know that the localized
one-magnon states are highly linearly dependent and that this will
also be the case for the corresponding multi-magnon states except for
states close to maximal packing. These results are corroborated by
computer-algebraic and numerical calculations for some finite
lattices ranging from $N=16$ to $N=256$. Thus the pyrochlore case
is only partly understood and seems to be the major challenge for
further research.

\section{General definitions\label{sec:D}}
We consider a spin system $(\Sigma, s,\hat{H})$ of $N=|\Sigma |$
spins with general spin quantum number
$s\in\{\frac{1}{2}, 1, \frac{3}{2}, \ldots \}$ and a
Heisenberg Hamiltonian
\begin{equation}\label{D1}
\hat{H}=\sum_{\mu,\nu\in\Sigma} J_{\mu\nu}
\left\{
\hat{S}_\mu^z \hat{S}_\nu^z+
\frac{1}{2}
\left(
\hat{S}_\mu^+ \hat{S}_\nu^- +\hat{S}_\mu^- \hat{S}_\nu^+
\right)
\right\}
-h \hat{S}^z
\;.
\end{equation}
where $h$ is the (dimensionless) magnetic field.
Since the $z$-component of the total spin
$\hat{S}^z=\sum_{\mu\in\Sigma}\hat{S}_\mu^z$
commutes with $\hat{H}$, its eigenvalue $M$ (the magnetic quantum number)
can be used to characterize the eigenstates of $\hat{H}$.
$\Sigma$ is assumed to be {\it connected},
that is, it cannot be divided into two subsystems
$\cal A$ and $\cal B$ such that $J_{\mu\nu}=0$
whenever $\mu\in{\cal A}$ and $\nu\in{\cal B}$.\\
The Hilbert space of the spin system is spanned by the basis of
product states $|m_1,\ldots,m_N\rangle$ which are simultaneous
eigenstates of $\hat{S}_\mu^z$ with eigenvalues $m_\mu$. Let
$|0\rangle = |s,s,s,\ldots\rangle$ denote the fully polarized
eigenstate of $\hat{H}$ (``magnon vacuum") and $|\mu\rangle =
(2s)^{-1/2} \hat{S}_\mu^-|0\rangle$ the $1$-magnon state localized
at spin site $\mu\in\Sigma$,
which will not be an eigenstate of $\hat{H}$.\\

The concept of $a$-LM states does not necessarily presuppose
the Heisenberg model, but also works with a
more general XXZ-model. For the XY-model the condition for
multi-magnon states can be relaxed. However,
all examples considered in this article will be Heisenberg
models and so we stick to this case in what follows.\\

Let $\cal G$ denote the {\it symmetry group} of $\Sigma$,
i.~e.~the group of permutations $\sigma$ of $\Sigma$ satisfying
$J_{\sigma\mu,\sigma\nu}=J_{\mu,\nu}$ for all $\mu,\nu\in\Sigma$.
The $1$-LM states to
be considered will be concentrated on a given subsystem
$L\subset\Sigma$ or its transforms $\sigma L,\; \sigma\in{\cal
G}$. In figures 1-4 the typical subsystems $L$ are indicated by
thick lines. A {\it localized $1$-magnon state with support} $L$
is an eigenstate of $\hat{H}$ with magnetic quantum number
$M=Ns-1$ of the form
\begin{equation}\label{D2}
\varphi=\sum_{\mu\in L} c_\mu |\mu\rangle,
\quad c_\mu\in\mathbb{C},\; c_\mu\neq 0\;.
\end{equation}
It follows that for $\sigma\in{\cal G}$ the state
\begin{equation}\label{D3}
\sigma\varphi=\sum_{\mu\in L} c_\mu |\sigma\mu\rangle=
\sum_{\nu\in \sigma L} c_{\sigma^{-1}\nu} |\nu\rangle
\end{equation}
will be a $1$-LM state with support $\sigma L$. Note that $\sigma L$ does not uniquely
determine $\sigma$,
since we may have $\sigma L=\tau L$ for $\sigma\neq\tau\in{\cal G}$.
Hence also the state (\ref{D3}) is not uniquely determined by $\sigma L$.
In order to fix the amplitudes of the states (\ref{D3}) we consider the subgroup
${\cal G}_L$ of all permutations of ${\cal G}$ leaving $L$ invariant. We choose a permutation from
each left coset of ${\cal G}/{\cal G}_L$ and denote the resulting set by
${\cal G}^L=\{\sigma_1,\ldots,\sigma_{\ell_1}\}$, where
\begin{equation}\label{D4}
|{\cal G}^L|=|{\cal G}/{\cal G}_L|=\ell_1\;.
\end{equation}
This yields the unique $1$-LM states with support $L^{(i)}\equiv \sigma_i L$
\begin{equation}\label{D5}
\varphi^{(i)}= \sigma_i\varphi=
\sum_{\nu\in \sigma_i L} c_{\sigma_i^{-1}\nu} |\nu\rangle
\equiv \sum_{\nu\in L^{(i)}} c_\nu^{(i)} |\nu\rangle,\;i=1,\ldots,\ell_1
\;.
\end{equation}
The set of these states will be denoted by ${\cal L}_1$ and the corresponding subsystems
$L^{(i)},\; i=1,\ldots \ell_1$ will be called {\it units}.
${\cal L}$ denotes the set of units.
By construction, ${\cal G}$ operates transitively on ${\cal L}$, i.~e.~all units are equivalent.
\\

Next we consider $a$-LM states.
Two different units $L^{(i)},\;L^{(j)}$ are called {\it overlapping} iff they contain
at least one common spin site, $L^{(i)}\cap L^{(j)}\neq\emptyset$, otherwise they are called
{\it disjoint}. Moreover, two different units $L^{(i)},\;L^{(j)}$ are called {\it connected}
iff there exist spin sites $\mu\in L^{(i)}$ and $\nu\in L^{(j)}$ such that $J_{\mu\nu}\neq 0$.
Otherwise, $L^{(i)},\;L^{(j)}$ are called {\it unconnected}.
Sometimes it will be convenient to identify the unit $L^{(i)}$ with its index $i$
and to write $i\sim j$ in case of connected units.\\

Let $1<a\le N$ be some integer and ${\cal S}=\{L^{i_1},\ldots,L^{i_a}\}\subset{\cal L}$
be a set of mutually unconnected units carrying $1$-LM states.
Further let $\mu_j\in\Sigma$ for $j=1,\ldots,a$ be different spin sites and denote by
$|\{\mu_1,\ldots,\mu_a\}\rangle$ the product state $|m_1,\ldots,m_N\rangle$ where
$m_{\mu_j}=s-1$ for $j=1,\ldots,a$ and $m_\mu=s$ else. As indicated by the notation, this state
will be invariant under permutations of the set $\{\mu_1,\ldots,\mu_a\}$.
Then
\begin{equation}\label{D6}
\Phi=\sum_{\mu_j\in L^{(i_j)}} c_{\mu_{1}}^{(i_1)}\ldots  c_{\mu_{a}}^{(i_a)}|\{\mu_1,\ldots,\mu_a\}\rangle
\end{equation}
will be an eigenvector of $\hat{H}$ with magnetic quantum number $M=Ns-a$. States of this kind will be called
localized $a$-magnon states ($a$-LM states). ${\cal L}_a$ denotes the set of all $a$-LM states.
The number $\ell_a$ of $a$-LM states equals the number of subsets
of mutually unconnected subsystems ${\cal S}\subset {\cal L}$. Hence
there exists a maximal number $a_{{\max}}$ of $a$-LM states such that
$\ell_a=0$ for $a>a_{\max}$. The sum
\begin{equation}\label{D7}
\ell\equiv \sum_{a=1}^{a_{\max}} \ell_a
\end{equation}
is the number of LM states.\\
If all units are pairwise unconnected, $a_{\max}=\ell_1$, see the
examples 1(a)(b)(d). In general, $a_{\max}\le \ell_1$. For the
XY-model, not considered in this article, the condition of
mutually unconnected units for $a$-LM states can be relaxed to the
condition of mutually disjoint units. Hence the XY-model admits
more $a$-LM states than the corresponding Heisenberg spin system
if the spin system contains disjoint, connected units, as in the
example of the checkerboard lattice (figure \ref{fig03}a).
\\

The following remarks apply to most of the examples considered
above but are not intended as additional assumptions for the
general theory to be outlined in the next sections. In the case of
spin lattices, the symmetry group $\cal G$ contains an abelian
subgroup $\cal T$ of translations. In this case it would be
natural to choose ${\cal G}^L=\{\sigma_1,\ldots,\sigma_{\ell_1}\}$
to consist of as many translations as possible. In fact, in all
examples mentioned above, except for the pyrochlore
(fig.~\ref{fig04}), it is possible to choose ${\cal G}^L=\cal T$.
This means that the translations operate transitively on the
lattice $\cal L$ of units or, equivalently, on the set ${\cal
L}_1$ of $1$-LM states. It would then be appropriate to denote the
$1$-LM states by $\varphi_{\bf r}$, where ${\bf r}$ denotes, for
example, the vector pointing to the midpoint of the corresponding
unit. Consequently, it is possible to decompose the linear span of
${\cal L}_1$ into irreducible representations of $\cal T$ which
are known to be one-dimensional and spanned by states of the form
$\psi_{\bf k}=\sum_{\bf r}\varphi_{\bf r}\exp(i {\bf k}\cdot {\bf
r})$ where ${\bf k}$ runs through the Brioullin zone of the
lattice. The states $\psi_{\bf k}$ are still ground states of the
Hamiltonian for the subspace $M=N s-1$ corresponding to a constant
ground state energy and hence form a so-called flat band. It is
obvious that the states $\psi_{\bf k}$ are linearly independent
iff the $\varphi_{\bf r}$ are so since both Gram matrices are
unitarily equivalent.\\
However, one has to be careful in the case of linear dependence of
the $1$-LM states. It is easily seen that also the space of linear
relations between $1$-LM states can be decomposed into vectors of
the form $\exp(i {\bf k}_0\cdot {\bf r})$ such that the
corresponding linear relations assumes the form $\psi_{{\bf
k}_0}=\sum_{\bf r}\varphi_{\bf r}\exp(i {\bf k}_0\cdot {\bf
r})=0$. Indeed, the states $\psi_{\bf k}$ are pairwise orthogonal
and span the same space as the $1$-LM states
which is impossible if all $\psi_{\bf k}\neq 0$.
Hence to $d$ independent linear relations between $1$-LM
states there correspond $d$ relations of the form
$\psi_{{\bf k}_0}=0$, i.~e.~$d$ ``holes" in the flat band.
The holes are ground states of the flat band which are not spanned by $1$-LM
states. We will come back to this question when dealing with concrete examples
in section \ref{sec:E}.\\
The pyrochlore lattice is more complicated and has to
be considered in more detail elsewhere.

\section{The Gram matrix\label{sec:G}}

Let $(\varphi_1,\ldots,\varphi_n),\;\varphi_i\in{\cal H}\text{ for }i=1,\ldots,n$
be a finite sequence of vectors in some Hilbert space ${\cal H}$
and let $G$ be the corresponding $n\times n$-matrix of all scalar products:
\begin{equation}\label{G1}
G_{ij}=\langle \varphi_i| \varphi_j\rangle,\; i,j=1,\ldots,n\;.
\end{equation}
$G$ is called the {\it Gram matrix} corresponding to the sequence of vectors. It is Hermitean and has
only non-negative eigenvalues, see \cite{Lan}. Moreover, the rank of $G$ equals the dimension of the linear span of
$(\varphi_1,\ldots,\varphi_n)$. Especially, $(\varphi_1,\ldots,\varphi_n)$ is linearly independent
iff $\det G > 0$ \cite{Lan}.\\
We will apply this criterion ({\it Gram criterion}) to sequences of $a$-LM states. In this case, we
will call the dimension of the null space of the Gram matrix of the $a$-LM states the {\it codimension}.
It thus equals the number of independent linear relations between $a$-LM states.
Without loss of generality we may assume that $G_{ii}=1$ for $i=1,\ldots, n$ in the sequel. If
$G_{ij}=\langle \varphi_i| \varphi_j\rangle =\delta_{ij}$, then the sequence $(\varphi_1,\ldots,\varphi_n)$
is obviously linearly independent. This case will be called the {\it orthogonal} case in the context of $a$-LM states.
If $G_{ij}=\langle \varphi_i| \varphi_j\rangle$ is sufficiently small for $i\neq j$ the Gram matrix will still
be invertible and the sequence will be linearly independent. For example, the following criterion easily
follows from Ger\v{s}gorin's theorem \cite{Lan} adapted to our problem:
\begin{lemma}\label{L1}
If the Gram matrix $G$ satisfies
\begin{equation}\label{G2}
\sum_{j=1,\;j\neq i}^n |G_{ij}|<1\text{ for all } i=1,\ldots,n
\end{equation}
then all eigenvalues of $G$ are strictly positive and $\{\varphi_1,\ldots,\varphi_n\}$ is linearly independent.
\end{lemma}

Next we will express the Gram matrix $G^{(a)}$ of $a$-LM states in terms of the Gram matrix $G^{(1)}$
of $1$-LM states. This will yield the result that the $a$-LM states are linearly independent if the $1$-LM states
are so. It will be sufficient to give the details only for $a=2$ and to leave the case $a>2$ to the reader.\\
We will use the following lemma from linear algebra:
\begin{lemma}\label{L2}
Let $A: {\cal H} \longrightarrow {\cal H}$ be a linear, positive semi-definite operator and ${\cal T}$ be
a subspace of ${\cal H}$ with the projector $P:{\cal H} \longrightarrow {\cal T}$ and the embedding
$P^\ast:{\cal T} \longrightarrow {\cal H}$.
\begin{enumerate}
\item[(1)] If $A$ is positive definite, then also its restriction $PAP^\ast$ will be positive definite.
\item[(2)] Let ${\cal N}$ be the null space of $A$, then ${\cal N}\cap {\cal T}$ will be the null space of $PAP^\ast$.
\end{enumerate}
\end{lemma}
{\bf Proof}:
Obviously, (2) implies (1).
For the proof of (2) we note that $A\psi=0$ and $\psi\in{\cal T}$ imply $PAP^\ast\psi=0$. Hence
${\cal N}\cap {\cal T}$ is a subspace of the null space of $PAP^\ast$. To show the other inclusion,
we assume that $\psi\in{\cal T}$ is in the null space of $PAP^\ast$. Hence $\langle \psi |A\psi\rangle = 0$.
Expanding $\psi$ into the eigenbasis of $A$ and utilizing that $A$ is positive semi-definite, we conclude
that $\psi\in{\cal N}$.
\hfill $\square$ \\

We now consider two $2$-LM states $\Phi^{(12)},\;\Phi^{(34)}$ supported by the pairwise unconnected
units $L^{(1)},\;L^{(2)}$
and $L^{(3)},\;L^{(4)}$, resp. They hence can be written as
\begin{eqnarray}\label{G3a}
\Phi^{(12)} &=&
\sum_{\scriptsize\begin{array}{l}\mu\in L^{(1)}\\ \nu\in L^{(2)}\end{array}}
c_\mu^{(1)} c_\nu^{(2)} |\{\mu,\nu\}\rangle \\ \label{G3b}
\Phi^{(34)} &=&
\sum_{\scriptsize\begin{array}{l}\kappa\in L^{(3)}\\ \lambda\in L^{(4)}\end{array}}
c_\kappa^{(3)} c_\lambda^{(4)} |\{\kappa,\lambda\}\rangle
\;.
\end{eqnarray}
The scalar product of these states is
\begin{eqnarray}\nonumber
G^{(2)}_{(12)\,(34)} &=&\langle \Phi^{(12)}|\Phi^{(34)}\rangle
\\ \label{G4}
&=&
\sum_{\mu,\nu,\kappa,\lambda}
\overline{c_\mu^{(1)} c_\nu^{(2)}}
c_\kappa^{(3)} c_\lambda^{(4)}
\langle \{\mu,\nu\}|\{\kappa,\lambda\}\rangle
\;.
\end{eqnarray}
The scalar product $\langle \{\mu,\nu\}|\{\kappa,\lambda\}\rangle$ is non-zero
only if $\{\mu,\nu\}=\{\kappa,\lambda\}$, that is $(\mu=\kappa \text{ and } \nu=\lambda)
\text{ or } (\mu=\lambda \text{ and } \nu=\kappa)$. In this case the scalar product equals $1$.
Hence
\begin{eqnarray}\nonumber
G^{(2)}_{(12)\,(34)} &=&
\sum_{\scriptsize\begin{array}{l}\mu\in L^{(1)}\cap L^{(3)}\\ \nu\in L^{(2)}\cap L^{(4)}\end{array}}
\overline{c_\mu^{(1)}}
c_\mu^{(3)}
\overline{c_\nu^{(2)}}
c_\nu^{(4)}
\\ \label{G5a}
&+&
\sum_{\scriptsize\begin{array}{l}\mu\in L^{(1)}\cap L^{(4)}\\ \nu\in L^{(2)}\cap L^{(3)}\end{array}}
\overline{c_\mu^{(1)}}
c_\mu^{(4)}
\overline{c_\nu^{(2)}}
c_\nu^{(3)}
\\ \label{G5b}
&=& G^{(1)}_{13}G^{(1)}_{24}+G^{(1)}_{14}G^{(1)}_{23}
\;.
\end{eqnarray}
If we ignore for a moment the condition of the units being unconnected, we could reformulate the equation
(\ref{G5b}) for general indices as
\begin{equation}\label{G5}
G^{(2)}_{(ij)\,(kl)} = G^{(1)}_{ik}G^{(1)}_{jl}+G^{(1)}_{il}G^{(1)}_{jk}
\;.
\end{equation}
This equation can be viewed as a statement saying
that $G^{(2)}$ is the restriction of $ G^{(1)}\otimes G^{(1)}$ to the symmetric subspace
of $\mathbb{C}^n \otimes \mathbb{C}^n$ spanned by the basis vectors
$(ij)\equiv (e_i\otimes e_j)_{\text{\scriptsize sym}}\equiv\frac{1}{\sqrt{2}}(e_i\otimes e_j+e_j\otimes e_i)$.
Here $e_i,\;i=1,\ldots,n$ denote the standard basis vectors of $\mathbb{C}^n$.
If $G^{(1)}$ is invertible, then also
$ G^{(1)}\otimes G^{(1)}$ is invertible,
since the eigenvalues of $ G^{(1)}\otimes G^{(1)}$ are the products of all pairs of
eigenvalues of $G^{(1)}$.
The same holds for the restriction of $ G^{(1)}\otimes G^{(1)}$
to the symmetric subspace of $\mathbb{C}^n \otimes \mathbb{C}^n$,
see lemma \ref{L2}(1).
Actually, $G^{(2)}$ is the further restriction of this matrix to
the subspace spanned by the basis vectors $(ij)$ such that the units $L^{(i)}$ and $L^{(j)}$
are unconnected, and hence $G^{(2)}$ is invertible too, invoking again lemma \ref{L2}(1).\\
The generalization to $a>2$ is straightforward. Let
$G^{(1)\otimes a}_{\text{\scriptsize sym}}$ be defined as the restriction of $ G^{(1)}\otimes G^{(1)}\otimes\ldots \otimes G^{(1)}$
to the totally symmetric
subspace $\mathbb{C}^{n\otimes a}_{\text{\scriptsize sym}}$ of
$\mathbb{C}^n \otimes \mathbb{C}^n \otimes\ldots \otimes \mathbb{C}^n$.
Then $G^{(a)}$ is the restriction of $G^{(1)\otimes a}_{\text{\scriptsize sym}}$ to an appropriate
subspace of $\mathbb{C}^{n\otimes a}_{\text{\scriptsize sym}}$
and hence, by lemma \ref{L2}(1), invertible if $G^{(1)}$ is so. In summary, we have proven
\begin{theorem}\label{T1}
If the set of $1$-LM states is linearly independent, then also the set of $a$-LM states is linearly independent
for all $a>1$.
\end{theorem}
In the general case where the linear span of $(\varphi_1,\ldots,\varphi_n)$ has the dimension $n-d$, or,
equivalently, where the Gram matrix has a $d$-dimensional null space, the above considerations can be
used to derive upper bounds for the codimension of the $a$-LM states. To this end we consider the eigenbasis of
$G^{(1)}$ instead of the standard basis of $\mathbb{C}^n$, such that the first $d$ basis vectors
belong to the eigenvalue $0$ of $G^{(1)}$.  The corresponding basis vectors of
$\mathbb{C}^{n\otimes a}_{\text{\scriptsize sym}}$ can be parametrized by sequences $(N_1,\ldots,N_n)$ of
``occupation numbers" satisfying  $N_i\ge 0$ and $\sum_{i=1}^n N_i=a$. This is familiar from Bose statistics.
Sequences of occupation numbers can equivalently be encoded as binary sequences of $n+a-1$ symbols
containing $n-1$ zeroes and $a$ ones. Such a sequence starts with $N_1$ ones, followed by a zero and
$N_2$ ones, and so on. There are exactly ${ n+a-1 \choose a}$ such sequences, which hence is the dimension
of $\mathbb{C}^{n\otimes a}_{\text{\scriptsize sym}}$. Those basis vectors of
$\mathbb{C}^{n\otimes a}_{\text{\scriptsize sym}}$
which do not involve any eigenvectors of
$G^{(1)}$ with the eigenvalue $0$ are characterized by occupation numbers with $N_1=\ldots=N_d=0$, or,
equivalently, by binary sequences commencing with $d$ zeroes. There are exactly
${ n+a-d-1 \choose a}$ such sequences, which is thus the rank of $G^{(1)\otimes a}_{\text{\scriptsize sym}}$.
Consequently, the null space ${\cal N}$ of $G^{(1)\otimes a}_{\text{\scriptsize sym}}$ has the dimension
${ n+a-1 \choose a}-{ n+a-d-1 \choose a}$.
Since $G^{(a)}$ is the restriction of $G^{(1)\otimes a}_{\text{\scriptsize sym}}$ to some appropriate subspace
${\cal T}$, ${ n+a-1 \choose a}-{ n+a-d-1 \choose a}$
is only an upper bound for the dimension of the null space ${\cal N}\cap{\cal T}$ of $G^{(a)}$, see
lemma \ref{L2}(2). Setting $n=\ell_1$ this yields the following theorem:
\begin{theorem}\label{T2}
If the codimension of $1$-LM states is $d$, then the codimension of $a$-LM states
is smaller or equal to ${ \ell_1+a-1\choose a}-{ \ell_1+a-d-1 \choose a}$.
\end{theorem}
Since the number $\ell_a$ of $a$-LM states is smaller or equal to ${\ell_1 \choose a}$
the estimate of theorem \ref{T2} will only be useful for small $a$ and not for
$a\approx a_{\max}$. Table 2 contains some examples illustrating theorem \ref{T2}.\\

We now turn to two other criteria concerning the codimension of $a$-LM states.
The first one is a criterion for linear independence.
\begin{theorem}\label{T3}
If there exists a spin site $\mu\in L$ which is not contained in any other unit $L^{(i)}$ (``isolated site"),
then the set of $a$-LM states is linearly independent for all $a\ge 1$.
\end{theorem}
Note that this theorem covers the examples of class $2$ (``isolated type") mentioned above.
For the {\bf proof} it suffices to consider the case $a=1$ by virtue of theorem \ref{T1}.\\
Assume $L=L^{(1)}$ and
\begin{equation}\label{G6}
0=\sum_{i=1}^{\ell_1} \lambda_i \varphi^{(i)} =\sum_{i=1}^{\ell_1} \lambda_i \sum_{\mu_i\in L^{(i)}}
c_{\mu_i}^{(i)}|\mu_i\rangle
\;,
\end{equation}
using (\ref{D5}).
By the assumption of the theorem, the state $|\mu\rangle$ occurs only once in the sum (\ref{G6}).
Hence $\lambda_1 c_\mu^{(1)} |\mu\rangle=0$, which gives $\lambda_1=0$,
since $c_\mu^{(1)}\neq 0$ by (\ref{D2}).
By symmetry arguments, the assumption of the
theorem holds for all $L^{(i)},\; i=1,\ldots,\ell_i$, hence all $\lambda_i$ vanish and the linear
independence of $(\varphi^{(1)},\ldots,\varphi^{(\ell_1)})$ is proven.
This completes the proof of theorem \ref{T3}.\\
\hfill $\square$ \\

The second criterion yields a bound for the codimension of $1$-LM states.
\begin{theorem}\label{T4}
Assume that every spin site is contained in some unit $L\in{\cal L}$.
Moreover, assume that there exists a unit $L\in{\cal L}$ with the property that
for each spin site $\mu\in L$ there exists exactly one other unit $L^{(i)}$ such that
$L\cap L^{(i)}=\{\mu\}$.\\
Then the codimension of ${\cal L}_1$ is less than or equal to one.
\end{theorem}
Note that this theorem covers all examples of the codimension one type, see fig.\ref{fig03}.
For the {\bf proof} we first show the following lemma:
\begin{lemma}\label{L3}
Under the assumptions of theorem \ref{T4}, the set ${\cal L}$ is ``connected" in the sense that
that for any two units $L^{(i)}$ and $L^{(j)}$ there exists a chain of pairs of overlapping units
which starts with $L^{(i)}$ and ends with $L^{(j)}$.
\end{lemma}
This lemma is readily proven by using the general assumption that the spin systems under consideration
are connected, see section \ref{sec:D}. Hence any two spin sites $\mu\in L^{(i)}$ and $\nu\in L^{(j)}$
can be connected by a chain of suitable spin sites $\mu_n$ such that $J_{\mu_n,\mu_{n+1}}\neq 0$.
By assumption, every $\mu_n$ is contained in some
unit $L^{(i_n)}$. It follows that $L^{(i_n)}$ and $L^{(i_{n+1})}$ are overlapping or identical units.
This concludes the proof of lemma \ref{L3}.\\
Returning to the proof of theorem \ref{T4} we
first note that, by the assumptions and symmetry arguments,
every spin site $\mu$ is contained in exactly two different units.
Let us consider again an equation of the form (\ref{G6}) and let $\mu$ be an arbitrary spin site.
By assumption, the state $|\mu\rangle$ occurs exactly twice in the sum (\ref{G6}), say
$\mu=\mu_i\in L^{(i)}$ and $\mu=\mu_j\in L^{(j)}$. It follows that
$\lambda_i c_{\mu_i}^{(i)}+\lambda_j c_{\mu_j}^{(j)}=0$. Since the amplitudes $c_{\mu_i}^{(i)}$
do not vanish, the ratio $\lambda_i/\lambda_j$ is uniquely fixed for every pair of overlapping
units $L^{(i)}$ and $L^{(j)}$ if not $\lambda_i=\lambda_j=0$.
By lemma \ref{L3}, all ratios $\lambda_i/\lambda_j$ are fixed or all $\lambda_i=0$. Hence
the space of vectors $\vec{\lambda}$ with coefficients $\lambda_i$ satisfying (\ref{G6})
is at most one-dimensional. This completes the proof of theorem \ref{T4}.
\hfill $\square$ \\

\section{Linear independence for codimension one type\label{sec:C}}

\begin{theorem}\label{T5}
Assume that all linear relations between $1$-LM states are multiples of
\begin{equation}\label{C2}
\sum_{i=1}^{\ell_1}\varphi^{(i)}=0\;.
\end{equation}
Then the set of $a$-LM states is linearly independent for $a>1$.
\end{theorem}
{\bf Proof}: We recall that the Gram matrix $G^{(a)}$ corresponding to the set of $a$-LM states is the
restriction of $G^{(1)\otimes a}$ to the subspace ${\cal T}^{(a)}$ of
$\mathbb{C}^{\ell_1\otimes a}_{\text{\scriptsize sym}}$
spanned by states of the form $(e_{j_1}\otimes\ldots\otimes e_{j_a} )_{\text{\scriptsize sym}}$,
where the $e_{j_n}$ are the standard basis vectors of $\mathbb{C}^{\ell_1}$ corresponding to
$1$-LM states $\varphi^{(j_n)}$ supported by pairwise unconnected units, see section \ref{sec:G}.
According to the assumption
of the theorem the null space of $G^{(1)}$ is
one-dimensional and spanned by the vector $\sum_{i=1}^{\ell_1}e_i$. Hence the null space
${\cal N}^{(a)}$ of
$G^{(1)\otimes a}$ is spanned by tensor products of the form
\begin{equation}\label{C4}
\sum_{k=1}^{\ell_1}e_{j_1}\otimes\ldots  \otimes e_k \otimes\ldots \otimes e_{j_{a-1}}
\;,
\end{equation}
where the position of the factor $e_k$ is fixed throughout the sum.
In view of lemma \ref{L2}(2) it remains
to show that
\begin{equation}\label{C5}
{\cal T}^{(a)}\cap {\cal N}^{(a)} = \{0\}
\;.
\end{equation}
To this end we assume that some linear combination of vectors in ${\cal N}^{(a)}$ of the form (\ref{C4})
lies in the subspace
${\cal T}^{(a)}$:
\begin{equation}\label{C6}
\sum_{j_1,\ldots,j_{a-1}} C_{j_1,\ldots,j_{a-1}}\sum_{k=1}^{\ell_1}
(e_{j_1}\otimes\ldots  \otimes e_k \otimes\ldots \otimes e_{j_{a-1}})_{\text{\scriptsize sym}}
\in{\cal T}^{(a)}
\;,
\end{equation}
where we have used the fact that ${\cal T}^{(a)}\subset\mathbb{C}^{\ell_1\otimes a}_{\text{\scriptsize sym}}$.
Hence the coefficients $C_{j_1,\ldots,j_{a-1}}$ will be invariant under permutations of their indices.
If we can show that all coefficients $C_{j_1,\ldots,j_{a-1}}$ vanish the linear independence of all
$a$-LM states follows and the proof of theorem \ref{T5} is done.
First we will rewrite the sum (\ref{C6}) as a linear combination of the standard basis vectors
$(e_{k_1}\otimes\ldots  \otimes e_{k_{a}})_{\text{\scriptsize sym}}$ of
$\mathbb{C}^{\ell_1\otimes a}_{\text{\scriptsize sym}}$. To this end it will be convenient to introduce some
special notation. \\

$\bf{k}$ is called a ``monotone multi-index of length $a$" iff ${\bf k}=(k_1,k_2,\ldots,k_a)$ such that
$k_1\le k_2\le\ldots\le k_a$. We will write
$ e_{\bf k}\equiv(e_{k_1}\otimes\ldots  \otimes e_{k_{a}})_{\text{\scriptsize sym}}$ and denote by
${\cal E}_a$ the set of all monotone multi-indexes of length $a$ such that at least two indices are equal.
Moreover we will denote by $\mbox{type}({\bf k})=(n_1,n_2,\ldots,n_a)$
the vector of numbers counting equal indices of ${\bf k}$.
For example, $\mbox{type}(1,1,2,2,2,5,8)=(2,3,1,1,0,0,0)$.
The symbol ${\bf j} \leftharpoonup {\bf k}$ will mean that ${\bf j}$ is obtained  by deleting
exactly one arbitrary index from $ {\bf k}$. For example, ${(1,1,2,3)} \leftharpoonup {(1,1,2,2,3)}$. The sum
$\sum_{{\bf j} \leftharpoonup {\bf k}}$ denotes the sum over all multi-indices obtained from $ {\bf k}$
by deleting exactly one index. It has $a$ terms if $a$ is the length of ${\bf k}$,
i.~e.~it will contain repeated terms if $ {\bf k}\in{\cal E}_a$.
For example, the sum $\sum_{{\bf j} \leftharpoonup {(1,1,2,3)}}$ runs over ${\bf j}=(1,2,3),(1,2,3),(1,1,3),(1,1,2)$.
\\
After this preparation we can rewrite the sum (\ref{C6}) in the form
\begin{equation}\label{C7}
\sum_{{\bf j} \leftharpoonup {\bf k}}
C_{\bf j}\; e_{\bf k}
\in{\cal T}^{(a)}
\;.
\end{equation}
Recalling the correspondence between indices and units it is obvious that
$e_{\bf k}\perp {\cal T}^{(a)}$
if $e_{\bf k}\in{\cal E}_a$,
since ${\cal T}^{(a)}$ is spanned by basis vectors $e_{\bf k}$ with even unconnected
indices (units). Hence
\begin{equation}\label{C8}
\sum_{{\bf j} \leftharpoonup {\bf k}}
C_{\bf j}\; =0 \quad\mbox{if}\quad
e_{\bf k}\in{\cal E}_a
\;.
\end{equation}
Our plan is to show that all $C_{\bf j}=0$ by downward induction over $n_1$ if
$\mbox{type}({\bf j})=(n_1,n_2,\ldots,n_{a-1})$.\\
$n_1=a-1$: Here ${\bf j}=(j,j,\ldots,j)$. Append one more $j$ to obtain ${\bf k}=(j,j,\ldots,j)\in{\cal E}_a$.
Hence
\begin{equation}\label{C9}
0=\sum_{{\bf i} \leftharpoonup {\bf k}}
C_{\bf i}\; = a\; C_{\bf j}
\quad \Rightarrow\quad C_{\bf j}=0
\;.
\end{equation}
$n_1+1 \rightarrow n_1$: Let $\mbox{type}({\bf j})=(n_1,n_2,\ldots,n_{a-1})$ and choose ${\bf k}$ by pre-pending
one more $j_1$ to ${\bf j}$. Hence $\mbox{type}({\bf k})=(n_1+1,n_2,\ldots,n_{a-1},0)$ and
${\bf k}\in{\cal E}_a$. It follows that
\begin{equation}\label{C10}
0=\sum_{{\bf i} \leftharpoonup {\bf k}}
C_{\bf i}\; =(n_1+1) C_{\bf j}
+ \sum_{\bf m}C_{\bf m}
\;,
\end{equation}
where the ${\bf m}$'s in the sum $\sum_{\bf m}C_{\bf m}$ denotes multi-indexes obtained from ${\bf k}$ by
deleting the next index of ${\bf k}$ different from $j_1$ and so on. Hence all ${\bf m}$ in this sum satisfy
$\mbox{type}({\bf m})=(n_1+1,\ldots)$ and thus $C_{\bf m}=0$ by the induction assumption. Then it follows from
(\ref{C10}) that $C_{\bf j}=0$.
This concludes the induction proof and the proof of theorem \ref{T5}.
\hfill $\square$ \\

\section{Special examples\label{sec:E}}
We consider the four types of examples mentioned in the introduction separately.
\subsection{Orthogonal type\label{sec:E1}}
For all four examples of this type the units are disjoint and any two different $1$-LM states are
orthogonal. The same property holds for $a$-LM states and hence all $a$-LM states are linearly
independent for $a\ge 1$.

\subsection{Isolated type\label{sec:E2}}
For all three examples of this type the units contain isolated spin sites, and hence the conditions of
theorem \ref{T3} are satisfied. Consequently, all $a$-LM states are linearly independent for $a\ge 1$.

\subsection{Codimension one type\label{sec:E3}}

The units of the kagom\'{e} lattice are hexagons and the amplitudes of the normalized $1$-LM states are the
alternating numbers $\pm \frac{1}{\sqrt{6}}$. We fix the amplitudes of one hexagon and obtain the amplitudes
of the other $1$-LM states by translation. Hence the amplitudes at one spin site of two $1$-LM states with
overlapping units have different signs and the corresponding scalar product gives $-\frac{1}{6}$. Since every
hexagon has exactly $6$ overlapping neighbors, the row sums of the Gram matrix $G^{(1)}$ excluding the diagonal are $-1$
and the Ger\v{s}gorin criterion of lemma \ref{L1} does not apply.
In fact, it is easy to see that the sum of all
$1$-LM states vanishes and hence the codimension of these states is at least one.
(Note that this relation between the $1$-LM states was given already in
Ref.~\cite{ZT05}.)\\
However, in the case of the checkerboard lattice (fig. 3a) the
amplitudes of $1$-LM states at the vertices of overlapping squares
are the same. Hence the linear relation assumes the form
$\sum_\mathbf{r} \pm \varphi_\mathbf{r}=0$, where the $1$-LM
states of every other row are multiplied by $-1$. According to the
discussion at the end of section \ref{sec:D} this corresponds to a
wave vector of $\mathbf{k}_0=(\pi,\pi)$, if the coordinate system
for the $\mathbf{k}$-vectors is oriented along the square's diagonals.\\
In order to show that the codimension is at most one we invoke
theorem \ref{T4}, since all three examples of codimension
one type satisfy the assumptions of this theorem.\\

Alternatively, to prove that the codimension is exactly one for
the kagom\'{e} lattice, we may use the fact that the non-diagonal
elements of $G^{(1)}$ are less or equal to $0$ and that the Gram
matrix is irreducible. This means that, even after arbitrary
permutations of the $1$-LM states, the Gram matrix does not have a
block structure with vanishing non-diagonal blocks which follows
from lemma \ref{L3}. Then the theorem of Frobenius-Perron
\cite{Lan} can be applied and proves that the lowest eigenvalue of
$G^{(1)}$, which is $0$, is non-degenerate.
The example of the star lattice (fig. 3b) is analogous.\\
\\

\begin{figure}
\begin{center}
\includegraphics[width=150mm]{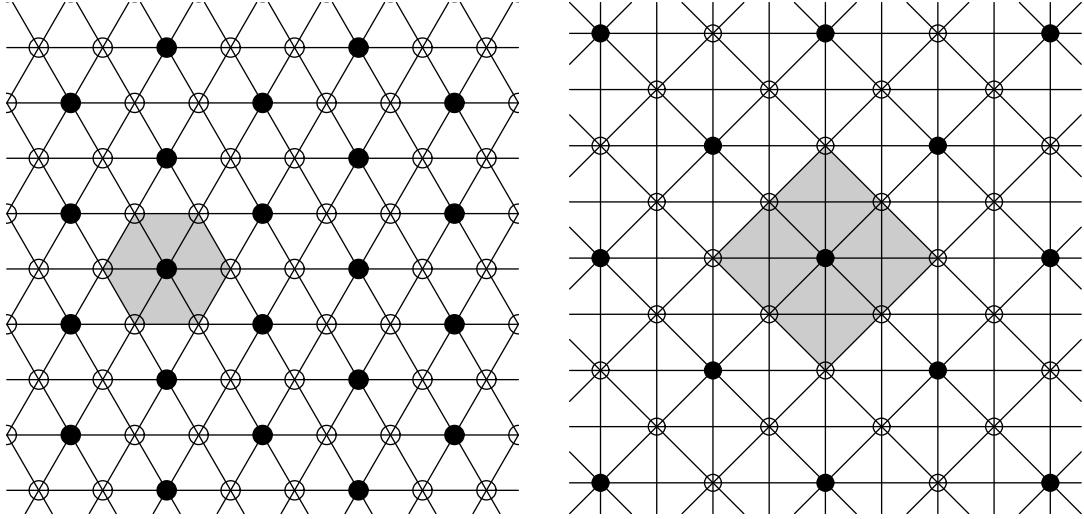}
\caption{\label{fig07}Left figure: The triangular lattice ${\cal L}$ of units (indicated by small circles)
for the kagom\'{e} lattice, showing an $a$-LM state with $a=a_{\max}= \frac{1}{3}\ell_1$. The occupied units
are indicated by filled circles.
For this state the lattice ${\cal L}$ can be filled
by hexagons drawn around each occupied unit. One of these hexagons is shown in the figure.\\
Right figure: The analogous figure for the checkerboard lattice. Here the lattice ${\cal L}$ of units
is a quadratic lattice with diagonals. The figure shows
an $a$-LM state with $a=a_{\max}= \frac{1}{4}\ell_1$. For this state the lattice ${\cal L}$ can be filled
by $4$-squares drawn around each occupied unit. One of these $4$-squares is shown in the figure.
}
\end{center}
\end{figure}

We note that two units of the kagom\'{e} lattice are disjoint iff they are unconnected.
The set of units ${\cal L}$ can be viewed as an undirected graph, the edges of which are formed
by pairs of connected units $i\sim j,\;i,j\in{\cal L}$.
For the kagom\'{e} and the star lattice case the graph of ${\cal L}$ is isomorphic to the triangular lattice.
\\
We have  the following
\begin{lemma}\label{L4}
For the kagom\'{e} lattice and the star lattice case the maximal number $a_{\max}$ of pairwise unconnected units
does not exceed $\frac{1}{3}$ of the number of all units $\ell_1$.
\end{lemma}
One can {\bf prove} lemma \ref{L4} geometrically by drawing a hexagon $H_i$ around each occupied
unit $i\in{\cal L}$ of an $a$-LM state. Due to the condition of pairwise disjointness, different hexagons
will not overlap. For the special state realizing $a = \frac{1}{3}\ell_1$, see figure \ref{fig07}, these
hexagons fill the whole triangular lattice ${\cal L}$. For finite versions of the kagom\'{e} lattice
we can only conclude $a_{\max}\le \frac{1}{3}\ell_1$,
since the filling of ${\cal L}$ with hexagons need not
be consistent with the periodic boundary conditions.
\hfill $\square$ \\

For the checkerboard lattice the graph of ${\cal L}$ is isomorphic to the square lattice with diagonals,
see figure \ref{fig07}. Each occupied unit is connected to $8$ unoccupied units which span a
$4$-square containing $4$ small squares. These large squares may fill the whole graph ${\cal L}$
if the size and the periodic boundary conditions are suitably chosen. Hence we have
\begin{lemma}\label{L4a}
For the checkerboard lattice the maximal number $a_{\max}$ of pairwise unconnected units
does not exceed $\frac{1}{4}$ of the number of all units $\ell_1$.
\end{lemma}
Note that for the checkerboard lattice with a maximal occupation satisfying $a_{\max}=\frac{1}{4}\ell_1$
there are two possible positions for each diagonal row of $4$-squares \cite{Z}. This explains the relatively
high degeneracy of the $a$-LM ground state for $a=a_{\max}$ which exceeds the corresponding
degeneracy in the case of the kagom\'{e} lattice or the star lattice.\\

For the kagom\'{e} and the star lattice the linear relation between $1$-LM states
assumes the form (\ref{C2}) and hence theorem \ref{T5} proves the linear
independence of all $a$-LM states for $a>1$. In contrast, the $1$-LM states of the checkerboard
lattice satisfy $\sum_i \pm \varphi^{(i)}=0$, as mentioned above.
However, after a suitable re-definition of the states $\varphi^{(i)}$ this sum
also can be brought into the form (\ref{C2}). More generally, this is possible for any
linear relation of the form
\begin{equation}\label{E1}
\psi_{\mathbf{k}_0} =\sum_\mathbf{r} e^{i\,\mathbf{k}_0\cdot\mathbf{r}}\varphi_\mathbf{r}=0
\;,
\end{equation}
see the remarks at the end of section \ref{sec:D}.\\

Nevertheless it will be instructive to check theorem \ref{T5} for
some finite lattices by computer-algebraic and numerical methods.
In table $1$ we present the corresponding results for the
checkerboard lattice. Related results showing that the degeneracy
of the ground state exceeds $\ell_a$ for some finite
kagom\'{e} lattices have been obtained by A.~Honecker \cite{AH}.\\
Since the $1$-LM states of the checkerboard are concentrated on open squares
(i.~e.~without diagonals) and each spin site
belongs to two open squares, we have necessarily $\ell_1=N/2$ in all examples of table $1$.
The corresponding
rank of the Gram matrix, or, equivalently, the number of linearly independent $1$-LM states
is $\ell_1-1$,
since the checkerboard lattice has codimension one. The dimension
of the ground state in the sector $a=1$ is always $\ell_1+1$,
since there are two ground states not spanned
by $1$-LM states: they correspond to the wave vector
$\mathbf{k}_0=(\pi,\pi)$ and
belong to the flat (FB) and the dispersive band (DB), resp.~, see the remarks at the end of section \ref{sec:D}.
The amplitudes of these additional states can be written as
\begin{eqnarray}\label{E31a}
a_{FB}(n_1,n_2)
&=&
\left\{
\begin{array}{l@{\mbox{ if }}l}
0 &  n_1+n_2 \mbox{ is odd}\\
(-1)^{n_1} & n_1+n_2 \mbox{ is even,}
\end{array}
\right.\\
a_{DB}(n_1,n_2)
&=&
\left\{
\begin{array}{l@{\mbox{ if }}l}
0 &  n_1+n_2 \mbox{ is even}\\
(-1)^{n_1} & n_1+n_2 \mbox{ is odd.}
\end{array}
\right.
\end{eqnarray}
Here $(n_1,n_2)\in\mathbb{Z}^2$ are the integer coordinates of the
lattice sites of the checkerboard. These states can be viewed as superpositions
of ``chain states", which have alternating amplitudes $\pm 1$ along
certain lines.\\

As mentioned above, each unit of the checkerboard lattice is
surrounded by $8$ units which are connected to the first one, if
$N$ is sufficiently large ($N>16$ in our examples).
Hence, if one unit is occupied by an $1$-LM state,
there remain $\ell_1-9$ free units to host another
$1$-LM state. Hence $\ell_2=\frac{1}{2}\ell_1(\ell_1-9)$ for $N>16$,
see table \ref{Table1}. One can similarly
argue for $a=3$: If two disconnected units are occupied,
there are usually $\ell_1-18$ units left for a third $1$-LM state.
However, if the first two units are close together,
only $17, 16$ or $15$ units are connected with these and
hence forbidden for the third $1$-LM state.
A detailed calculation yields the formula
$\ell_3=\frac{1}{3!}\ell_1(\ell_1^2-27\ell_1+194)$
which is satisfied for our examples except for the smallest lattices with
$N = 16$ and $N=36$, see table \ref{Table1}.\\

In all cases where we have explicitly checked the linear
independence of $a$-LM states, $a>1$, for finite lattices, thereby
confirming theorem \ref{T5}, the number for the rank is listed in
table \ref{Table1}. The rank is always strictly less than the
dimension of the ground state (DGS). However, the ratio
$\mbox{DGS}/\mbox{rank}$ seems to approach $1$ if $N$ increases.
This would justify to count only the number of $a$-LM states as
ground states in the
thermodynamic limit $N\rightarrow\infty$.\\

\begin{table}
\caption{\label{Table1}Computer-algebraical and numerical results for various finite checkerboard lattices
(fig. \ref{fig03}a) differing in size $N$. The spin quantum number is always $s=\frac{1}{2}$.
The periodic boundary conditions (PBC) are chosen with edge vectors parallel to the edges of the unit square (P)
or with both edge vectors rotated by $45^\circ$ (R).
The number $2A$ characterizes the length of the lattice in one dimension, hence $N=(2A)^2$ in case P.
$\ell_a$ denotes the number of $a$-LM states and ``Rank" the number
of linearly independent $a$-LM states. ``DGS" denotes the degeneracy of the true ground state
which was determined numerically. Empty entries correspond to numbers which have not been calculated.
}
\vspace{5mm}
\begin{indented}
\item[]\begin{tabular}{@{}llllll}
\hline
 PBC & $N$ & $2A$ & $\ell_a$ & Rank & DGS \\
\br
P& $16$ & $4$ & $\ell_1=8$ & $7$ & $9$\\
& && $\ell_2=4$ & $4$ & $13$\\
& && $\ell_3=0$ &  &$1$\\
\mr
P& $36$& $6$ & $\ell_1=18$ & $17$ & $19$\\
& && $\ell_2=81$ & $81$ & $118$  \\
 &&& $\ell_3=84$ & $84$  & $250$\\
 &&& $\ell_4=18$ & $18$  & $83$\\
&& &$\ell_5=0$ &   & $1$\\
\mr
P& $64$& $8$ & $\ell_1=32$ & $31$ & $33$\\
&&& $\ell_2=368$ & $368$& $433$ \\
&&& $\ell_3=1888$ & & $2833$ \\
&&& $\ell_4=4392$ & &  \\
&& &$\ell_5=4224$& &  \\
\mr
P&  $100$& $10$ & $\ell_1=50$ & $49$ & $51$\\
&&& $\ell_2=1025$ &$1025$ & $1126$\\
&&& $\ell_3=11200$ & & $14026$ \\
&&& $\ell_4=71150$ & &  \\
\mr
P& $144$& $12$ & $\ell_1=72$ & $71$ & $73$\\
&&& $\ell_2=2268$ & $2268$ & $2413$\\
&&& $\ell_3=41208$ & &  \\
\mr
P& $256$& $16$ & $\ell_1=128$ & $127$ & $129$\\
&&& $\ell_2=7616$ & $7616$ & $7873$\\
&&& $\ell_3=279936$ & &  \\
\mr
R&$32$& $4$ & $\ell_1=16$ & $15$ & $17$\\
&&& $\ell_2=56$ & $56$ & $89$\\
&&& $\ell_3=48$ &$48$ & $137$ \\
&&& $\ell_4=12$ &$12$ & $31$ \\
&& &$\ell_5=0$ & & $3$ \\
\br
\end{tabular}
\end{indented}
\end{table}

\subsection{Higher codimension type, pyrochlore lattice\label{sec:E4}}

The infinite pyrochlore lattice can be obtained from a single tetrahedron by
the following construction.
Starting with
a tetrahedron with vertices at
$(0,0,0),(0,1,1),(1,0,1),(1,1,0)$, one performs inversions
about all vertices.
In the next step the lattice points obtained by all possible inversions
about the new vertices are added, and so on, ad infinitum.
The analogous definition for two dimensions and the equilateral triangle yields the kagom\'{e} lattice, hence
the pyrochlore lattice can be viewed as the three-dimensional analogue of the kagom\'{e} lattice.\\
The finite versions of the pyrochlore lattice are obtained by assuming appropriate periodic boundary conditions (PBC).
As in other cases, there are various possibilities for the subgroups of translations which define the PBC. These
subgroups are generated by three vectors called ``edge vectors"\cite{SSRB}.
In our examples we have chosen as edge vectors either even multiples $2A$ of the vectors
$(0,1,1),(1,0,1),(1,1,0)$ (tetrahedral PBC) or even multiples $2A$ of the vectors
$(0,0,1),(0,0,1),(1,0,0)$ (cubic PBC).
\\
The units of the pyrochlore lattice which support the $1$-LM states are hexagons, see figure \ref{fig04}.
Each lattice site is contained in $6$ hexagons, and each hexagon contains, of course, $6$ lattice sites.
Hence the number of $1$-LM states, which equals the number of hexagons, is $\ell_1 = N$.
For sufficiently large $N$ ($N>32$ in our examples) each hexagon has a common edge with $6$ other hexagons
and a common vertex (without having a common edge) with $18$ other hexagons. Hence,
if one unit is occupied by an $1$-LM state there remain $N-(6+18+1)=N-25$ units free for a second
$1$-LM state. Consequently, $\ell_2=\frac{1}{2}N(N-25)$ for $N>32$, see table \ref{Table2}.\\

\begin{figure}
\begin{center}
\includegraphics[width=150mm]{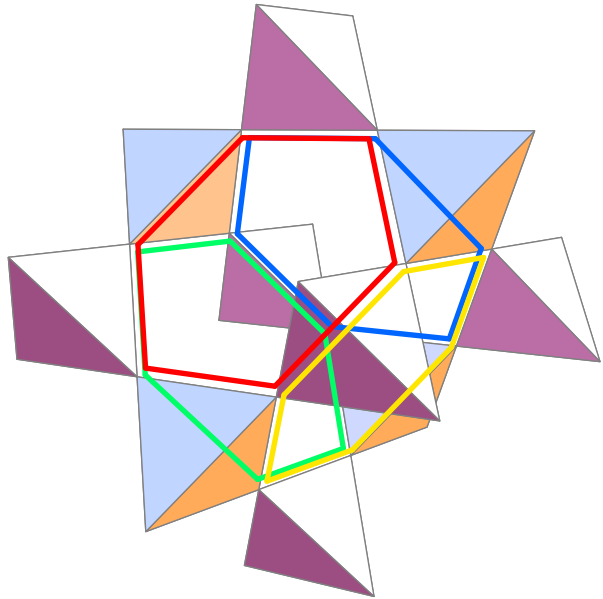}
\caption{\label{fig08}Four localized magnon states on a pyrochlore
lattice supported by hexagons at the faces of a super-tetrahedron.
The sum of these four states vanishes, which explains almost
completely the linear dependence of $1$-LM states. }
\end{center}
\end{figure}

For our purposes
these finite pyrochlore lattices are interesting,
since they provide the first examples of lattices
admitting $a$-LM states with codimension higher than one. This has been shown for
$a=1$ and $a=2$ and can be shown to hold also for larger $a$, see table $2$ and the last paragraph
of this section.\\
That the codimension of the $1$-LM states on pyrochlore lattices
${\cal P}$ exceeds $1$ follows already from the fact that its
restriction to a plane spanned by $3$ vertices of any tetrahedron
in ${\cal P}$ will be isomorphic to the kagom\'{e} lattice. For
every such plane the sum of the $1$-LM states concentrated on
hexagons lying in that plane vanishes. Hence we obtain a
considerable number of linear dependencies
among the $1$-LM states of ${\cal P}$. \\
Another kind of linear dependence among $1$-LM states \cite{HH,Z2}
can be understood by the vanishing of the sum of four $1$-LM
states (with suitably chosen signs) supported by four hexagons
lying on the faces of a super-tetrahedron, see figure \ref{fig08},
and spanning a so-called truncated tetrahedron, or
super-tetrahedron, see \cite{MW}. These super-tetrahedra have
edges which are three times as long as the edges of the tetrahedra
forming the pyrochlore lattice. Each super-tetrahedron contains
$4$ hexagons, and each hexagon lies in two super-tetrahedra. Hence
there are exactly $N/2$ super-tetrahedra, and the same number of
linear relations between $1$-LM states. However, these linear
relations are not independent; they only span a space of dimension
$N/2-1$. In the examples considered there are always
exactly three additional linear relations which are not due to super-tetrahedra.\\
The rank of the Gram matrix $G^{(1)}$, which equals the number of
linearly independent $1$-LM states is thus always equal to
$N/2-2$. The degeneracy of the ground state $\mbox{DGS}$ in the
sector $a=1$ amounts to $N/2+1$, see table $2$. This corresponds
to three additional ground states which are not spanned by $1$-LM
states and belong to the wave vector $\mathbf{k}=\mathbf{0}$ and
three different bands: two flat bands and one dispersive band.
Alternatively, the additional ground states can be thought as
three chain states along three linearly independent directions of
the tetrahedron. For $a>1$ table $2$ confirms the inequalities
$\ell_a > \mbox{rank} > \mbox{LB}$, where $\mbox{LB}$ is the lower
bound of $\mbox{rank}$ derived from theorem \ref{T2}, and
$\mbox{DGS} > \mbox{rank}$. We expect that finite size effects are
especially strong for the considered $3$-dimensional pyrochlore
lattices of the size $16\le N\le 256$ due to the restrictions of
the periodic boundary conditions. Nevertheless, the results of
table $2$ do not contradict the conjecture that
$\mbox{DGS} \approx \mbox{rank}$ in the limit $N\rightarrow \infty$.\\

The existence of ``localized linear dependencies" between $1$-LM
states has an interesting consequence for the codimension of
$a$-LM states for $a>1$. Contrary to the situation of theorem
\ref{T5} it is now possible to construct non-trivial linear
relations between $a$-LM states in the following way. We will use
again the notation of section \ref{sec:C} and write the linear
dependence of $4$ states localized at a super-tetrahedron $T$ in
the form $\sum_{k=1}^4 e_k=0$. Further we consider $1$-LM states
with indices $j_1,\ldots, j_{a-1}$ supported by hexagons which are
pairwise disjoint and disjoint to $T$. Then the vector
\begin{equation}\label{E41}
\sum_{k=1}^4
(e_{j_1}\otimes\ldots  \otimes e_k \otimes\ldots \otimes e_{j_{a-1}})_{\text{\scriptsize sym}}
\end{equation}
lies in the null space of $G^{(a)}$ and thus corresponds to a linear relation between $a$-LM states.
This argument will be valid for all $a$ except if $a$ approaches $a_{\text{\scriptsize max}}$.
Then it may happen that there don't exist $a-1$ hexagons with the required properties.
Hence a certain part of the codimension of $a$-LM states can be explained by
localized linear dependencies between $1$-LM states. This yields an upper bound for the
rank of $G^{(a)}$ which we have indicated in table \ref{Table2}
in the case where it could be calculated.\\

\begin{table}
\caption{\label{Table2}Computer-algebraical and numerical results for various pyrochlore lattices (fig. \ref{fig04})
differing in size $N$ and periodic boundary conditions (PBC). The spin quantum number is always $s=\frac{1}{2}$.
The number $2A$ characterizing the periodicity is explained in the text.
$\ell_a$ denotes the number of $a$-LM states and ``Rank" the number
of linearly independent $a$-LM states. ``LB" denotes the lower bound for the rank (if positive) obtained by
theorem \ref{L2}. ``UB" denotes the upper bound for the rank obtained by
localized linear dependencies between $1$-LM states, see section \ref{sec:E4}.
``DGS" denotes the degeneracy of the true ground state
which was determined numerically. Empty entries correspond to numbers which have not been calculated.
}
\vspace{5mm}
\begin{indented}
\item[]\begin{tabular}{@{}llllllll}
\hline
PBC & $2A$ & $N$ & $\ell_a$ & Rank & LB & UB & DGS \\
\br
tetrahedral & $4$ & $32$ & $\ell_1=32$ & $14$ &&& $17$\\
 &&& $\ell_2=160$ & $72$ && $104$& $89$\\
& && $\ell_3=96$ & $94$ & &  &$137$\\
& && $\ell_4=24$ & $24$ & & & $31$\\
\mr
tetrahedral &$6$& $108$ & $\ell_1=108$ & $52$ &&& $55$\\
& && $\ell_2=4482$ & $1252$ & & & $1324$\\
& &&  &  & &&$18244$\\
\mr
tetrahedral &$8$& $256$ & $\ell_1=256$ & $126$ &&& $129$\\
 &&& $\ell_2=29568$ & & $4673$ && $7873$\\
\mr
cubic & $4$& $16$ & $\ell_1=16$ & $6$ & &&$9$\\
&&& $\ell_2=0$ & & & &$13$\\
\mr
cubic &$8$& $128$ & $\ell_1=128$ & $62$ && &$65$\\
&&& $\ell_2=6592$ & $1825$ & $1289$ &&$1889$\\
\br
\end{tabular}
\end{indented}
\end{table}

\section*{Acknowledgement}
We thank O.~Derzhko, A.~Honecker and J.~Schnack for intensive
discussions concerning the subject of this article and A.~Honecker for a critical
reading of the manuscript.
The work was supported by the DFG.
For the exact diagonalization J\"org Schulenburg's {\it spinpack} was used.

\end{document}